\documentclass[prb,twocolumn,superscriptaddress]{revtex4}
\usepackage{graphicx}% Include figure files
\usepackage{dcolumn}% Align table columns on decimal point
\usepackage{bm}% bold math
\usepackage[latin1]{inputenc}
\usepackage{amsfonts}
\usepackage{amssymb}
\usepackage{amsmath}

\begin{document}
%%%%%%%%%%%%%%%%%%%%%%%%%%%%%%%%%%%%%%%%%%%%%

\title{Acceptor and donor impurities in GaN nanocrystals}

%%%%%%%%%%%%%%%%%%%%%%%%%%%%%%%%%%%%%%%%%%%%%
\author{C. Echeverr\'ia-Arrondo}
\author{J. P\'erez-Conde}
\affiliation{Departamento de F\'isica, Universidad P\'ublica de Navarra, E-31006, Pamplona, Spain}
\author{A. K. Bhattacharjee}
\affiliation{Laboratoire de Physique des Solides, UMR du CNRS, Universit\'e Paris-Sud, F-91405, Orsay, France}
\date{\today}
%%%%%%%%%%%%%%%%%%%%%%%%%%%%%%%%%%%%%%%%%%%%%

\begin{abstract}

%%%%%%%%%%%%%%%%%%%%%%%%%%%%%%%%%%%%%%%%%%%%%
We investigate acceptor and donor states in GaN nanocrystals doped with
a single substitutional impurity. Quantum dots (QD's) of 
zinc-blende structure and spherical shape are considered with the radius ranging from 4.5 to 67.7~{\AA}. The size-dependent energy spectra are calculated within the $sp^3d^5s^{\ast}$ tight-binding model, which yields a good agreement with the confinement-induced blue shifts observed in undoped QD's. The computed binding energy is strongly enhanced with respect to the experimental bulk value  when the dopant is placed at the center of the smallest QD's. It decreases with
increasing QD size following a scaling law that extrapolates to the bulk limit. In order to estimate the degree of localization of the bound carriers we analyze their wave functions and average radii. The resulting physical picture points to a highly localized acceptor hole, mostly distributed over the nearest-neighbor anion shell, and a much more extended donor electron. We also study off-center impurities in intermediate-size QD's. The acceptor binding energy is approximately independent of the dopant position unless it is placed within a surface shell of thickness of the order of the bulk Bohr radius, where the ionization energy abruptly drops. On the contrary, the donor binding energy gradually decreases as the impurity is moved away from the center toward the QD surface.
\end{abstract}
\maketitle
%%%%%%%%%%%%%%%%%%%%%%%%%%%%%%%%%%%%%%%%%%%%

\section{INTRODUCTION}

%%%%%%%%%%%%%%%%%%%%%%%%%%%%%%%%%%%%%%%%%%%%%
Group-III nitride semiconductors such as AlN, GaN and InN have drawn considerable attention in recent years because they generate efficient luminescence and their band gaps include the entire visible spectrum and a part of the ultraviolet (UV). As a result of intense research on these compounds, visible light-emitting diodes and UV laser diodes are available nowadays, opening the way to further developments in optoelectronics, specially in optical storage technology. These materials are also advantageous in high-power and high-temperature electronics. We focus on GaN, since it is the most outstanding nitride of this group, widely used and researched.\cite{orton,vurgaftman,morko,reshchikov} 

The great interest in semiconductor quantum dots (QD's) is based on the possibility of tailoring their electrooptical properties through the control of size and doping. Nanocrystalline\cite{leppert98,micic99,preschilla00} and self-assembled\cite{APL69-4096,APL80-3937,APL83-984,PRB74-75305,JAP102-24311,PRB75-125306} QD's of GaN have been synthesized and show the confinement-induced blue shift. Recently, Eu-doped GaN nanocrystals (NC's) yielding a bright and narrow Eu$^{3+}$ ion fluorescence line have been shown to be useful as an efficient marker of important biological processes.\cite{fluorescence} However, to our knowledge, no experimental study of donor- or acceptor-doped GaN QD's has yet been reported. Theoretically, spherical NC's containing a substitutional impurity have been investigated  previously.\cite{einevoll,chuu,porras,jap74,zhu,deng,ferreira,jap81,yang,janis,ranjan,jap90,movilla,perez,pssb} Both donors and acceptors have been studied within a variational approach (VA)\cite{deng} and the effective-mass approximation (EMA).\cite{chuu,porras,jap74,zhu,ferreira,jap81,yang,janis,ranjan,jap90,movilla,pssb} Acceptors have also been treated in the atomistic framework: Effective bond-orbital\cite{einevoll} and tight-binding (TB)\cite{perez} models.

The electronic structure of a doped QD is highly sensitive to the position of the impurity atom. But most of the theoretical studies reported so far have focused on on-center impurities. The impurity position dependence has been investigated only for donors in the continuous models, VA and EMA, which are known to be inadequate for the electronic structure of small-size QD's and highly localized acceptor states even in the bulk. In the present work we report a TB model calculation of the electronic structure for both on- and off-center donor and acceptor impurities in GaN NC's of diameter up to 13.5~nm. A preliminary study reported previously\cite{perez} was limited to only on-center acceptor impurities in small-size NC's up to 4.5~nm in diameter. Also, it was based on the simple semi-empirical $sp^3s^{\ast}$ model of Vogl \textit{et al}.\cite{vogl} In contrast, the calculations presented here are based on the extended $sp^3d^5s^{\ast}$ model of Jancu \textit{et al}.,\cite{beltram} which provides a more satisfactory description of the band structure in the bulk. For the acceptor impurity we focus on Mg as an example. It substitutes for a Ga cation in the zinc-blende lattice. The hole binding energy in the bulk is known to be $236$~meV.\cite{przybylinska} For the donor we consider O, which replaces a N anion yielding an electron binding energy of $33$~meV in the bulk.\cite{reshchikov} The paper is organized as follows: In Section II we present an outline of the theoretical model; in Sections III and IV we show the binding energy and average radius of the bound carrier as a function of the dopant location and the QD size; in Section V we conclude by summarizing the main results.  
%%%%%%%%%%%%%%%%%%%%%%%%%%%%%%%%%%%%%%%%%%%%%

\section{THEORY}

%%%%%%%%%%%%%%%%%%%%%%%%%%%%%%%%%%%%%%%%%%%%%
The GaN host material is described by a semi-empirical $sp^3d^5s^{\ast}$ TB Hamiltonian which includes the spin-orbit interaction, and was first introduced by 
Jancu \textit{et al.}\cite{beltram}:
\begin{equation}
H_{TB}=\sum_{ij}\sum_{\mu\nu}h^{ij}_{\mu\nu}c_{i\mu}^\dagger c_{j\nu}+\sum_{i}\sum_{\mu\nu}\lambda_i<\mu|l \cdot s|\nu>c_{i\mu}^\dagger c_{i\nu},
\end{equation}
where \textit{i} and \textit{j} denote the atomic sites, $\mu$ and $\nu$ the
single-electron orbitals, and $h^{ij}_{\mu\nu}$ the hopping and on-site energies; $\lambda_i$ is a constant standing for the spin-orbit interaction. The tight-binding parameters account for the electronic structure of bulk GaN.\cite{jancu}

We need to take into consideration the dangling bonds on the surface of the quasi-spherical crystallite, since they introduce states within the band gap. The simplest way to move
these states several eV's away from the band gap region consists in binding every dangling bond to a hydrogen atom;\cite{perez110} the Ga-H and N-H distances are taken as one half the respective covalent bond lengths. This simple passivation procedure is more detailed in a previous work\cite{perez110} and references therein.

The effective Bohr radius $a_B^a$ of the hole bound to an acceptor in bulk GaN 
is typically small: In the case of Mg (binding energy $E_b^a=236$ meV) it is 4.5~{\AA}. The Bohr radius $a_B^d$ of the electron
bound to a donor is much larger: $a_B^d= 27.7$~{\AA} for O
($E_b^d=33$ meV). These values of the Bohr radii are obtained in the hydrogenic
model: $a_B=\hbar/\sqrt{2E_b(\rm{bulk})\,m^{\ast}}$, where the effective mass is 0.8~$\rm{m}_0$ for the bound hole\cite{perez} and 0.15~$\rm{m}_0$ for the bound electron.\cite{vurgaftman} Since the donor Bohr radius is over six times larger than that of the acceptor, we need to construct large enough QD's to follow the confinement effects on donors. The resulting Hamiltonian matrices are huge and sparse. For example, our largest QD consists of $\sim113,000$ atoms and gives a matrix of size over $2\cdot10^6\times2\cdot10^6$. Furthermore, the new (lower) symmetries introduced by the off-center impurities are difficult to handle within a group-theoretical framework, as in Ref.~\onlinecite{perez}. Hence, we resort to Lanczos algorithms\cite{cullum,matrix,klimeck} to calculate the few dopant levels we are interested in. The numerical libraries that we use are freely avaliable at the Arpack site.\cite{arpack}

The impurity is modelled by a screened Coulomb potential:  
\begin{equation}
U(r_i,R)= \left\{ \begin{array}{lll}
              \pm\frac{e^2}{\varepsilon(r_i,R)r_i}&\quad\rm{if}& r_i\neq0\\
	U_0&\quad\rm{if}&r_i=0
              \end{array}
    \right.,
\end{equation}
where $r_i$ is the distance between the lattice site \textit{i} and the impurity, \textit{R} the NC radius, and $\varepsilon(r_i,R)$ the dielectric function within the crystal. Here $U_0$ stands for the central cell potential which obviously cannot be described in terms of the long-range Coulomb potential that represents the impurity atom in its crystalline environment as an effective point charge. In the absence of any precise theoretical model, as usual, we shall treat $U_0$ as a phenomenological parameter. Position-dependent dielectric functions have been proposed by Resta\cite{resta} and by Franceschetti and Troparevsky.\cite{franceschetti} The position dependence is, however, restricted to a small volume around the dopant, typically smaller than the distance between two nearest-neighbor atoms,\cite{resta} beyond which the permitivity recovers the long-distance behavior. When restricted to QD's that means simply a value dependent on the QD size. We then take a size-dependent dielectric function given by the simplified Penn model\cite{tsu}:
\begin{equation}
\varepsilon(R)=1+\frac{\varepsilon_{B}-1}{1+(\xi/R)^\gamma}.
\end{equation}
The parameters entering this model are $\gamma=2$, the material-dependent $\xi=\pi E_F/k_F E_g=7.43$, by using $E_g=3.3$~eV and $E_F=16$~eV (Ref.~\onlinecite{orton}), and the bulk dielectric constant\cite{madelung} $\varepsilon_B(0)=8.9$. 

The acceptor and donor binding energies are respectively defined as\cite{kohn,perez}
\begin{equation}
\label{eqn:hl}
\begin{array}{l}
E_b^a=E_a-E_{\text{HOMO}},\\
E_b^d=E_{\text{LUMO}}-E_d,
\end{array}
\end{equation}
where $E_{\text{HOMO}}(R)$ is the energy of the valence band-edge state, the so-called highest occupied molecular orbital (HOMO), and $E_{\text{LUMO}}(R)$ the energy of the conduction band-edge state, the lowest unoccupied molecular orbital (LUMO), in the undoped NC. $E_a(R)$ and $E_d(R)$ correspond to the same states, but in the doped NC.

We also compute the average radius of the bound carrier, $\bar{r}(R)$, which is related to the effective Bohr radius at large $R$:
\begin{equation}
\label{eqn:aver}
\bar{r}(R\rightarrow\infty)=\frac{3a_B}{2}\varpropto \frac{1}{\sqrt{E_b(\rm{bulk})}}.
\end{equation}
%%%%%%%%%%%%%%%%%%%%%%%%%%%%%%%%%%%%%%%%%%%%%

\section{RESULTS: IMPURITY AT THE NANOCRYSTAL CENTER}

\begin{figure}[!ht]
\includegraphics[scale=0.3]{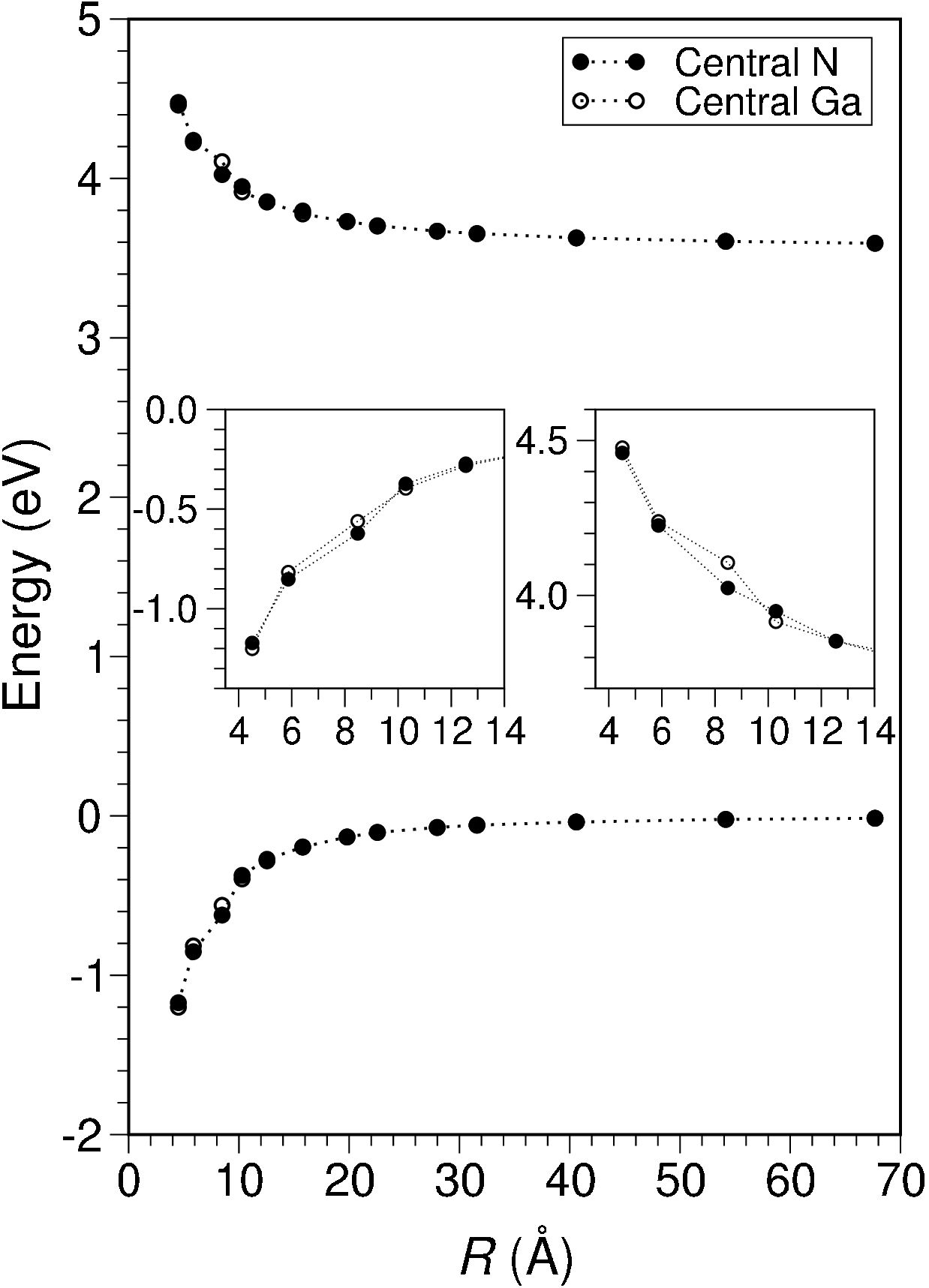}
\caption{\label{fig:fig1}Highest valence band (HOMO) and lowest conduction band
(LUMO) levels in Ga- and N-centered undoped NC's (open and closed symbols, respectively). In the insets the results for the smallest NC's are displayed. The dotted lines are drawn to guide the eye.}
\end{figure}

%%%%%%%%%%%%%%%%%%%%%%%%%%%%%%%%%%%%%%%%%%%%%
We first study the electronic structure of undoped QD's. The calculated energy levels $E_{\text{HOMO}}(R)$ and $E_{\text{LUMO}}(R)$ are plotted against the NC radius in Fig.~\ref{fig:fig1}. They are the reference levels for calculating the impurity binding energies (see Eq.~(\ref{eqn:hl})). Note the dramatic increase of the band gap with decreasing QD size. This confinement-induced blue shift is in accord with the reported spectroscopic studies of GaN NC's.\cite{leppert98,micic99,APL80-3937} Quantitatively, we obtain a satisfactory agreement with experiment: As an example, the calculated gap for NC's of 30~{\AA} diameter is roughly the same as the experimental value $3.65$~eV reported in Ref.~\onlinecite{micic99}. It is interesting to note that, in the full size range, the HOMO and LUMO degeneracies we obtain are compatible with the $\Gamma_8$ and $\Gamma_6$ representations, respectively, in accord with the symmetry analysis of Ref.~\onlinecite{perez}.

Another feature illustrated in Fig.~\ref{fig:fig1} is a comparison of the electronic structure in two different crystallizations: Ga- and N-centered NC's. As can be seen in the figure, the difference between them is negligible for $R>15$~{\AA}. Even in smaller NC's (see the insets) the small fluctuations do not seem to be significant. This behavior allows us to restrict our considerations to cation-centered NC's for an acceptor impurity and anion-centered NC's for a donor impurity even in the case of off-center impurities (see below) without any loss of generality.
%--------------------------------------------------------------------
%	Figura 1-Homo and Lumo levels
%--------------------------------------------------------------------
%--------------------------------------------------------------------

\begin{figure}[!ht]
\includegraphics[scale=0.3]{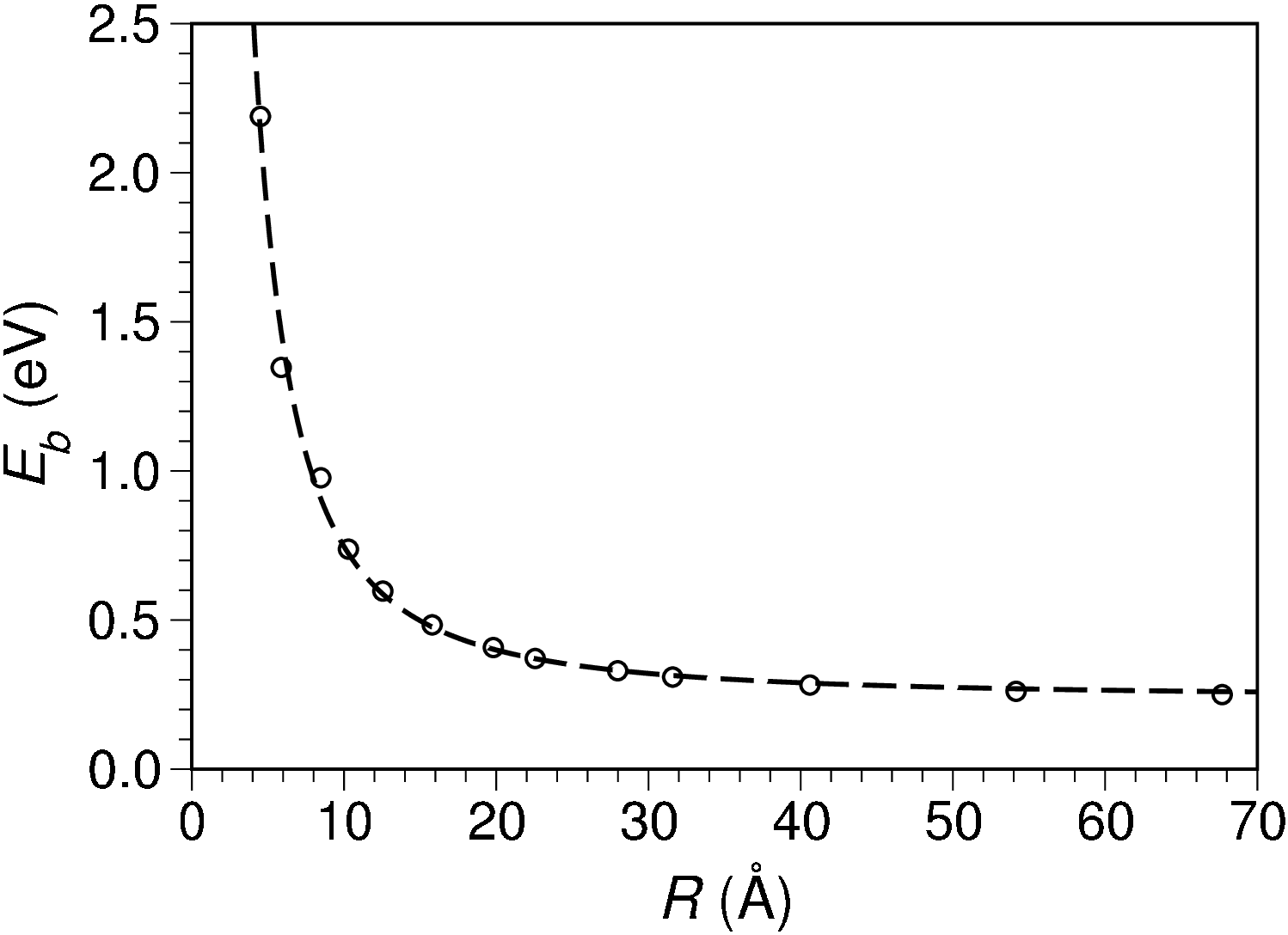}
\caption{\label{fig:fig2}NC size-dependent acceptor binding energy for $U_0=11$~eV and the fitting curve $E_b=0.239+23.149\,R^{-1.659}$ (dashed line).}
\end{figure}

\begin{figure}[!ht]
\includegraphics[scale=0.3]{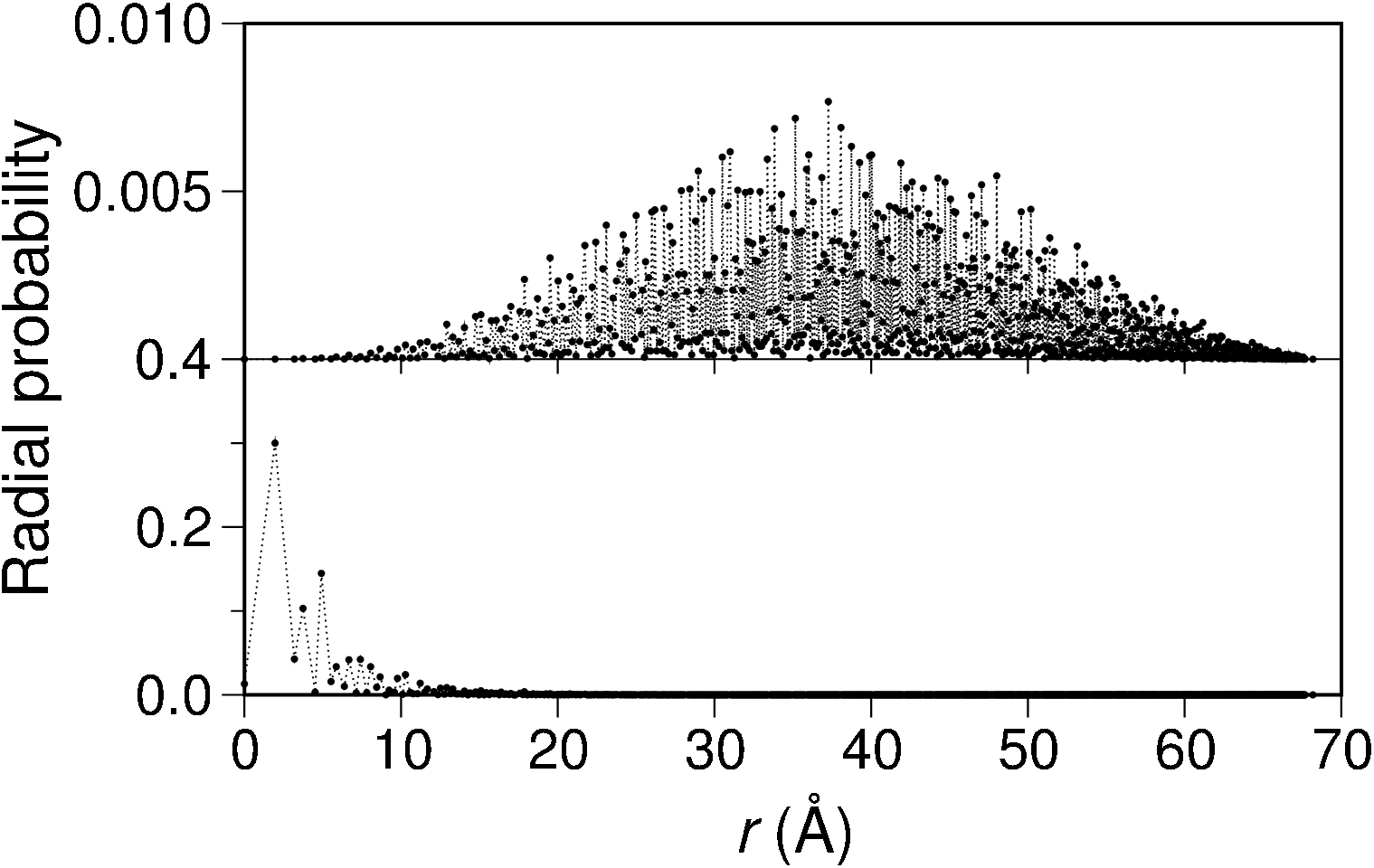}
\caption{\label{fig:fig3}Radial probability distribution of the HOMO in an undoped NC of 68~{\AA} radius (upper part), and of the ground-state acceptor hole in the same NC doped with a Mg atom (lower part). \textit{r} is the radial distance from the QD center. Note that the two probability scales are different. Dotted lines guide the eye.}
\end{figure}

\begin{figure}[!ht]
\includegraphics[scale=0.3]{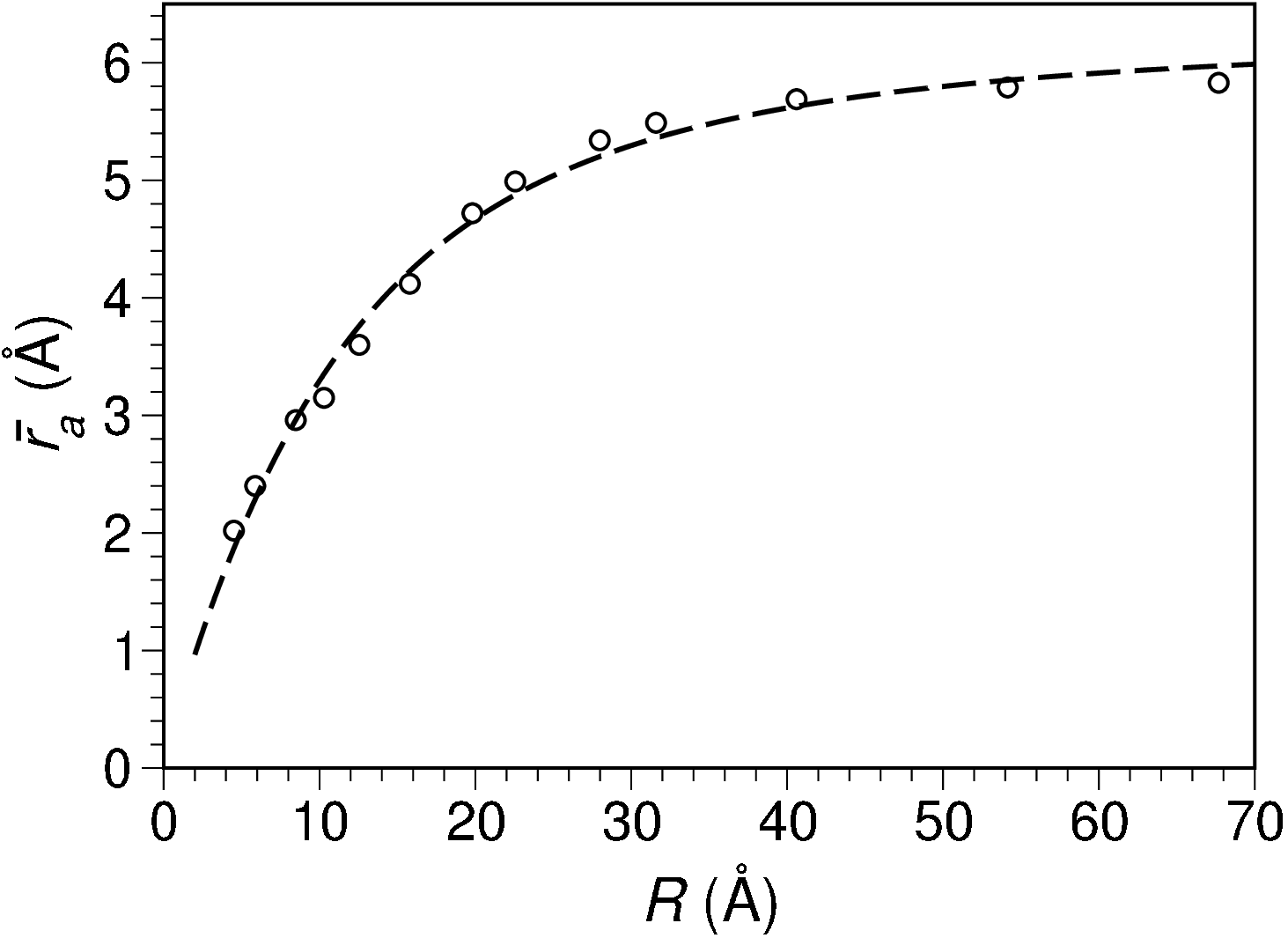}
\caption{\label{fig:fig4}QD size dependence of the average radius of the acceptor hole in the ground state and the fitting curve $\bar{r}_a=3.035/\sqrt{0.235+31.685\,R^{-1.713}}$ (dashed line).}
\end{figure}

We next discuss the impurity states starting with the acceptor impurity. We calculate the HOMO in NC's with a Mg atom replacing the Ga at the center and deduce the corresponding hole binding energy $E_b^a$ as a function of the NC radius for different values of the parameter $U_0$. The results are found to fit the simple expression (dropping the superscripts for simplicity)
\begin{equation}
\label{eqn:simple-exp}
E_b(R)=\epsilon_b +\textit{A}\cdot\textit{R}^{-\beta}, 
\end{equation} 
as shown in Fig.~\ref{fig:fig2} for $U_0=11$~eV. Note that this particular value of $U_0$ has been chosen for the Mg impurity in GaN, because the corresponding asymptotic limit  $E_b(R\rightarrow\infty)=\epsilon_b=0.239$~eV approximately reproduces the experimental bulk binding energy (0.236 eV). Complementary results concerning the QD acceptor states presented in Figs.~\ref{fig:fig3}, \ref{fig:fig4} and \ref{fig:fig8} correspond to the same value of $U_0$. The fitting parameters $\epsilon_b$, \textit{A} and $\beta$, of course, depend on $U_0$. They may also depend on the QD size range. Generally speaking, the size ranges can be distinguished in terms of two
different regimes of confinement: $R>a_B$ (weak confinement) and $R<a_B$ (strong confinement). In the strong-confinement regime the usual image of the impurity center as a hydrogen-like entity is no longer valid. The asymptotic value $E_b$(bulk) obviously corresponds to $\epsilon_b$ in the weak-confinement regime. Since the radii of all the investigated NC's are larger than the Mg acceptor Bohr radius, all of them belong to the weak-confinement regime in this case, and a single set of fitting parameters is sufficient as shown in Fig.~2.  
%--------------------------------------------------------------------
%	Figura 2-Acceptor binding energy
%--------------------------------------------------------------------
%--------------------------------------------------------------------

The degree of localization of the bound carriers is characterized through the analysis of wave functions and average radii. In Fig.~\ref{fig:fig3} we show the shell-wise radial probability distribution of the HOMO in an undoped QD of 68~{\AA} radius in the upper panel. The ground-state of the hole in the same QD when doped with a Mg atom is shown in the lower panel. Both are band-edge wave functions, remarkably different from each other: The HOMO is spread over the whole NC with a maximum at $r\simeq R/2$, while the bound hole is essentially concentrated on the five nearest-neighbor atomic shells. This highly localized character of the acceptor hole in the ground state remains unchanged over the whole set of NC's, as illustrated by the small size dependence of the average radius shown in Fig.~4, in agreement with experiment in the bulk. The magnetic resonance spectra in Mg-doped GaN layers are consistent with the same description of the hole, mainly distributed over the four nearest-neighbor atoms surrounding the impurity.\cite{kaufmann} The average radius of the hole orbit plotted as a function of the NC radius in Fig.~\ref{fig:fig4} is found to fit the curve
\begin{equation}
\label{eqn:radio}
\bar{r}_a(R)=\frac{c_1}{\sqrt{E_b(R)}}=\frac{c_1}{\sqrt{\epsilon_b+\textit{c}_2\cdot \textit{R}^{-\textit{c}_3}}},
\end{equation}
where $c_1$, $c_2$ and $c_3$ are the fitting parameters. From (\ref{eqn:aver}), this expression adequately reproduces the EMA bulk radius at large R: $\bar{r}_a(R\rightarrow\infty)=6.3$~{\AA}, which is indeed close to $3a_B^a/2 = 6.7$~{\AA}. 
%--------------------------------------------------------------------
%	Figura 3- Probability distributions
%--------------------------------------------------------------------
%--------------------------------------------------------------------
%	Figura 4-Average radius of the bound hole
%--------------------------------------------------------------------
%--------------------------------------------------------------------

\begin{figure}[!ht]
\includegraphics[scale=0.3]{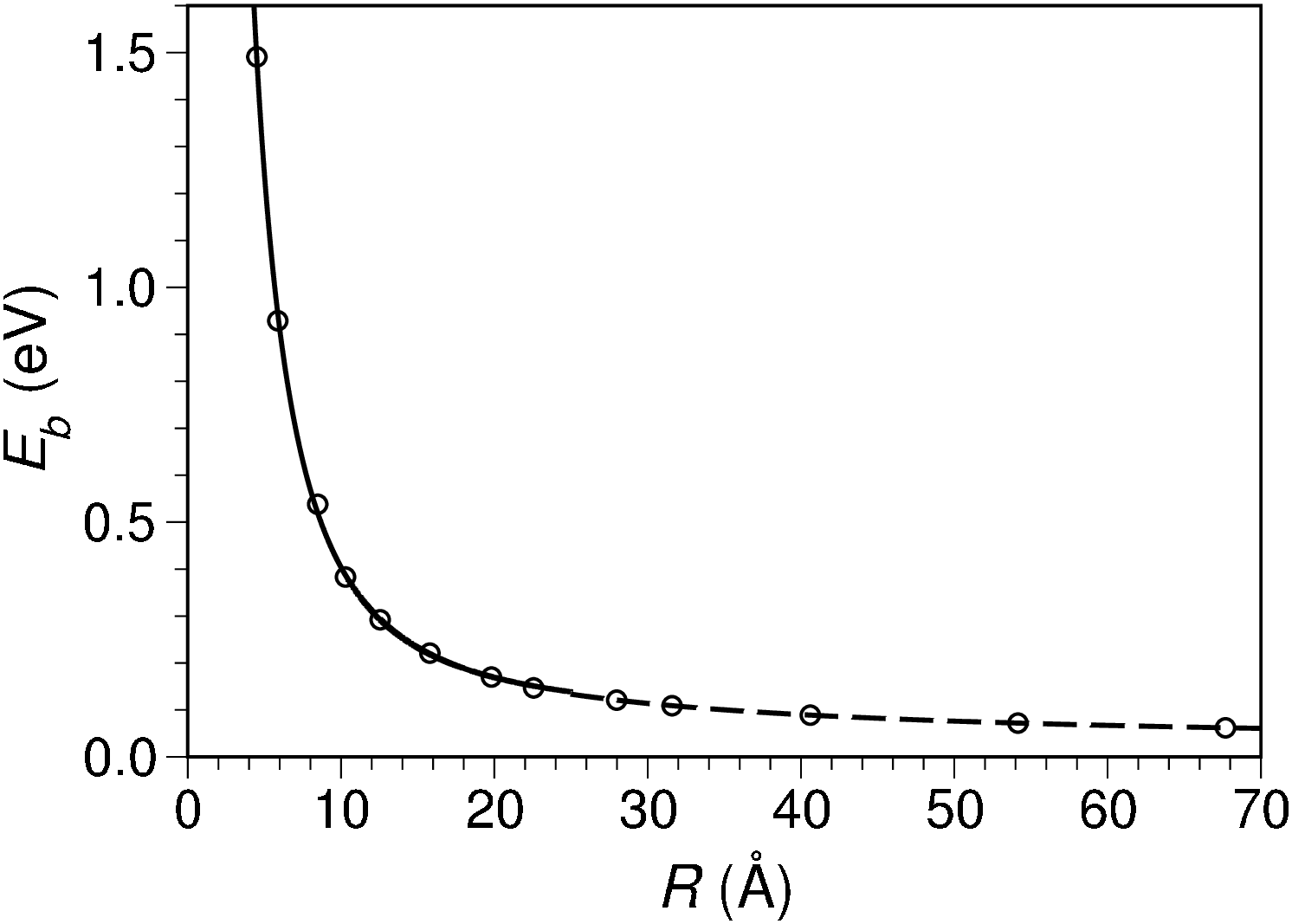}
\caption{\label{fig:fig5}QD size-dependent donor binding energy for $U_0=-2$~eV and the fitting curves: $E_b=0.078+22.386\,R^{-1.837}$ in the strong-confinement regime (solid line) and $E_b=0.027+3.860\,R^{-1.115}$ in the weak-confinement (dashed line).}
\end{figure}

\begin{figure}[!ht]
\includegraphics[scale=0.3]{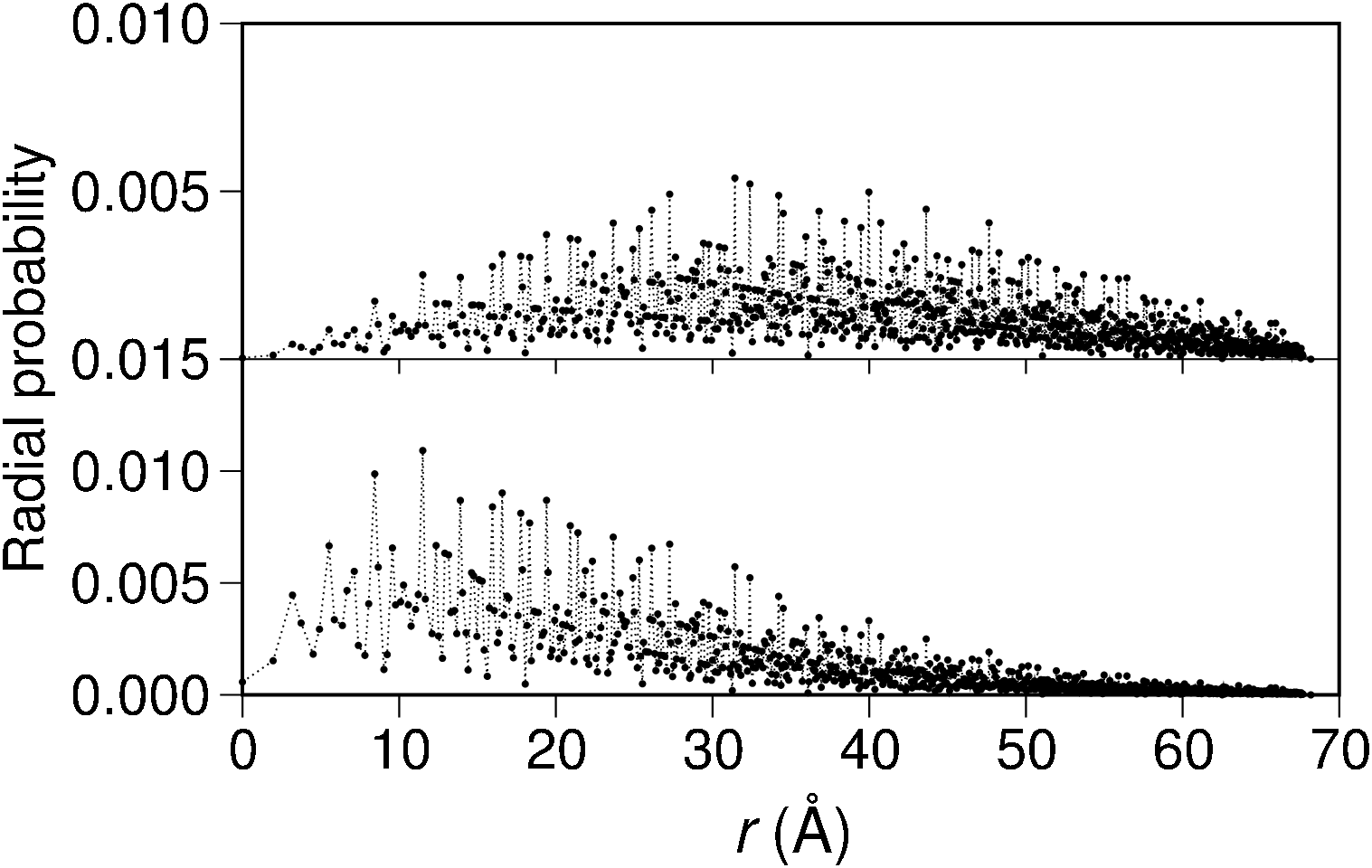}
\caption{\label{fig:fig6}Radial probability distribution of the LUMO in an undoped NC of 68~{\AA} radius (upper part) and that of the ground-state donor electron in the same NC when doped with an O atom (lower part). Note that the two probability scales are different.}
\end{figure}

\begin{figure}[!ht]
\includegraphics[scale=0.3]{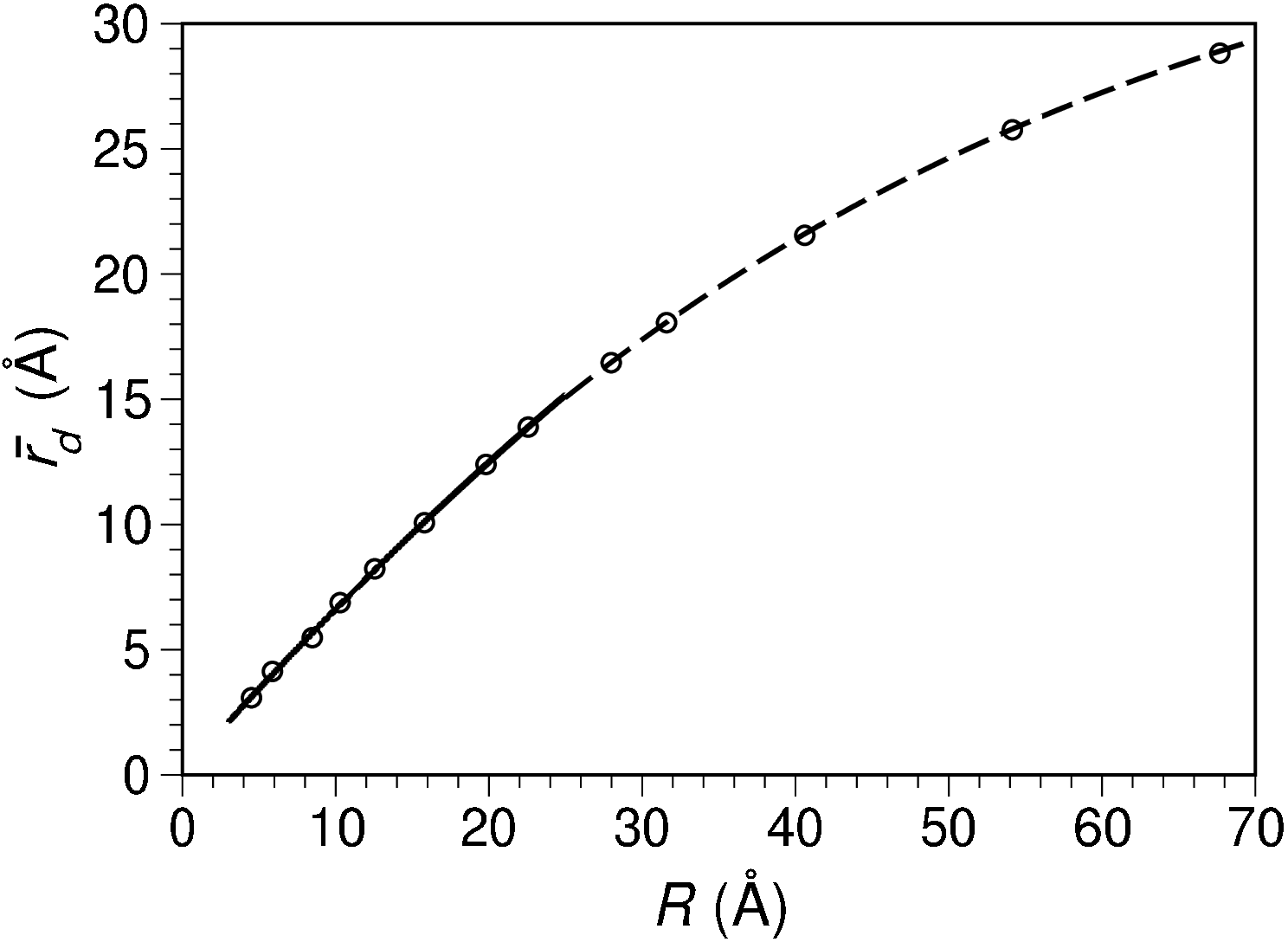}
\caption{\label{fig:fig7}QD size-dependent average radius of the donor electron in the ground state and the fitting curves: $\bar{r}_d=6.697/\sqrt{0.021+81.871\,R^{-1.912}}$ in the strong-confinement regime (solid line) and $\bar{r}_d=7.279/\sqrt{0.032+81.495\,R^{-1.865}}$ in the weak-confinement (dashed line).}
\end{figure}

\begin{figure}[!ht]
\centerline{}
\includegraphics[scale=0.28]{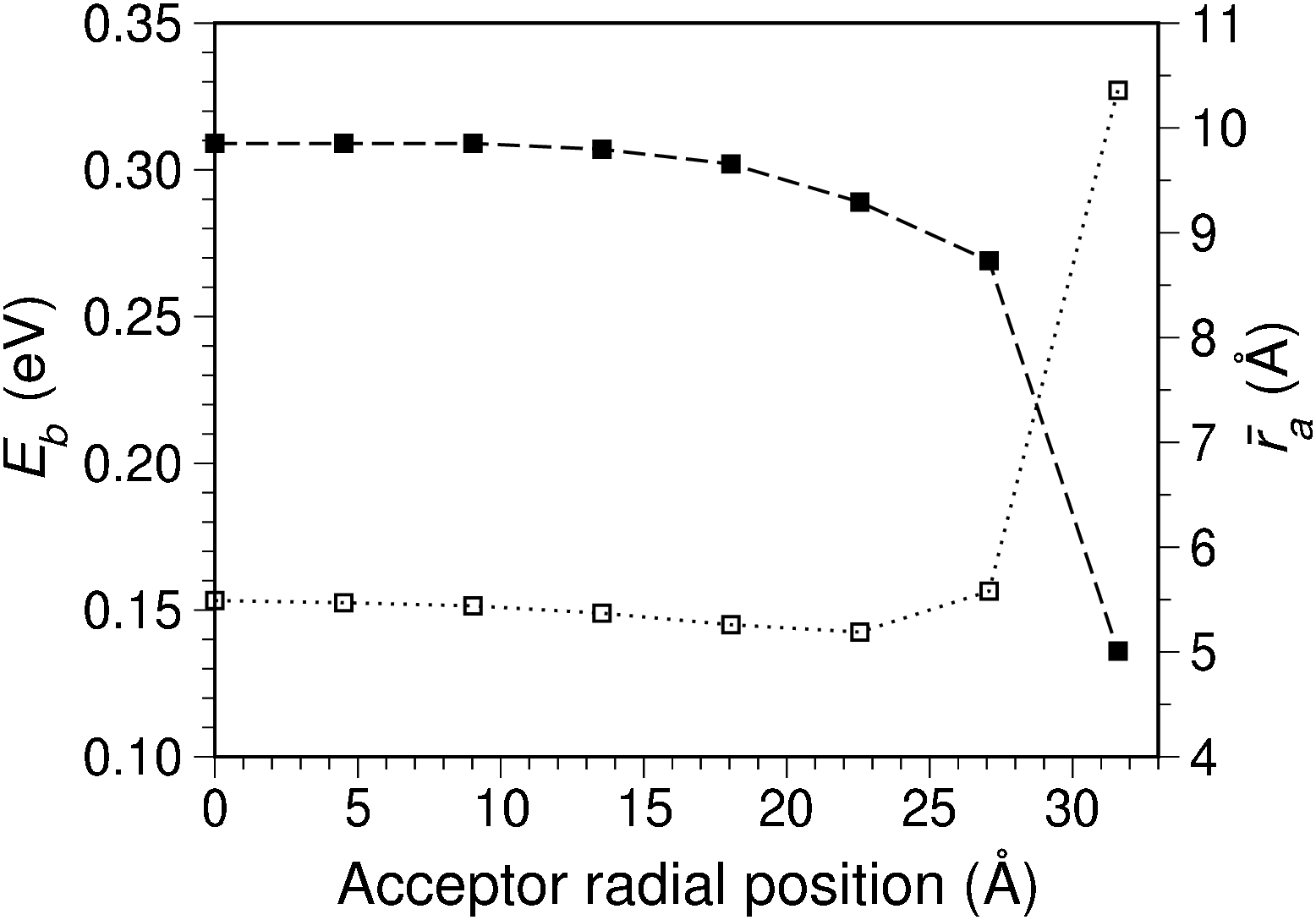}
\caption{\label{fig:fig8}Binding energy and average radius of the hole ground
state as a function of the acceptor position in a QD of radius 31.6~{\AA} (closed and open symbols, respectively). The dashed and dotted lines are drawn to guide the eye.}
\end{figure}

The ground state of the donor impurity (O) is investigated in a similar way. In order to obtain the value of $U_0$ suitable for the O impurity in GaN we computed the binding energies for different values of $U_0$ and fit their size dependence to the expression (\ref{eqn:simple-exp}). Two distinct sets of fitting parameters are obtained for the two confinement regimes. In Fig.~\ref{fig:fig5} we show the results for $U_0=-2$ eV: The binding energy is strongly enhanced in the smallest NC's and tends to 0.027 eV in the bulk limit, which is close to the experimental value $E_b^d(\rm{bulk})=0.033$~eV. This value of $U_0$ is then used in further analysis of the donor states: Figs.~6, 7 and 9. In Fig.~\ref{fig:fig6} we show the probability distribution of the LUMO in an undoped QD of 68~{\AA} radius in the upper panel. That of the ground-state donor electron in the same QD when doped with an O atom is shown in the lower panel. They are the same band-edge states, but very different from each other, because of the carrier localization in the latter. From a comparison of Fig.~\ref{fig:fig3} with Fig.~\ref{fig:fig6} we see that the donor electron is much more spread out than the acceptor hole, regardless of the crystal size. The average radius of the bound electron presented in Fig.~\ref{fig:fig7} confirms this difference; it is remarkably larger than that of the bound hole in Fig.~\ref{fig:fig4} for all the NC's. The fitting curve 
$\bar{r}_d(R)$ in the weak-confinement regime
tends to 40.7~{\AA} for $R\rightarrow\infty$, in good agreement with the EMA radius in bulk, $3a_B^d/2=41.6$~{\AA}. 

%--------------------------------------------------------------------
%	Figura 5-Donor binding energy
%--------------------------------------------------------------------
%--------------------------------------------------------------------
%	Figura 6- Probability distribution
%--------------------------------------------------------------------
%--------------------------------------------------------------------
%	Figura 7-Average radius of the bound electron
%--------------------------------------------------------------------
%--------------------------------------------------------------------
% \newpage
%%%%%%%%%%%%%%%%%%%%%%%%%%%%%%%%%%%%%%%%%%%%%

\section{RESULTS: OFF-CENTER IMPURITY}

%%%%%%%%%%%%%%%%%%%%%%%%%%%%%%%%%%%%%%%%%%%%%
We also compute off-center impurity states. As an example we focus on an intermediate-size NC of radius 31.6~{\AA}, slightly larger than the donor Bohr
radius ($\simeq28$~{\AA}). The impurity position dependence of the acceptor ground state is shown in Fig.~8, where the binding energy and the average radius are plotted against the distance of the impurity atom from the NC center. Notice the slow evolution until the acceptor approaches the QD surface, where dramatic changes occur: The binding energy drops and the average radius rises rather abruptly. This occurs when the Mg atom is placed within a surface layer of thickness
$\sim4.5$~{\AA} of the order of the acceptor Bohr radius. This behavior is related to the strongly localized nature of the bound hole state.

\begin{figure}[!ht]
\centerline{}
\includegraphics[scale=0.28]{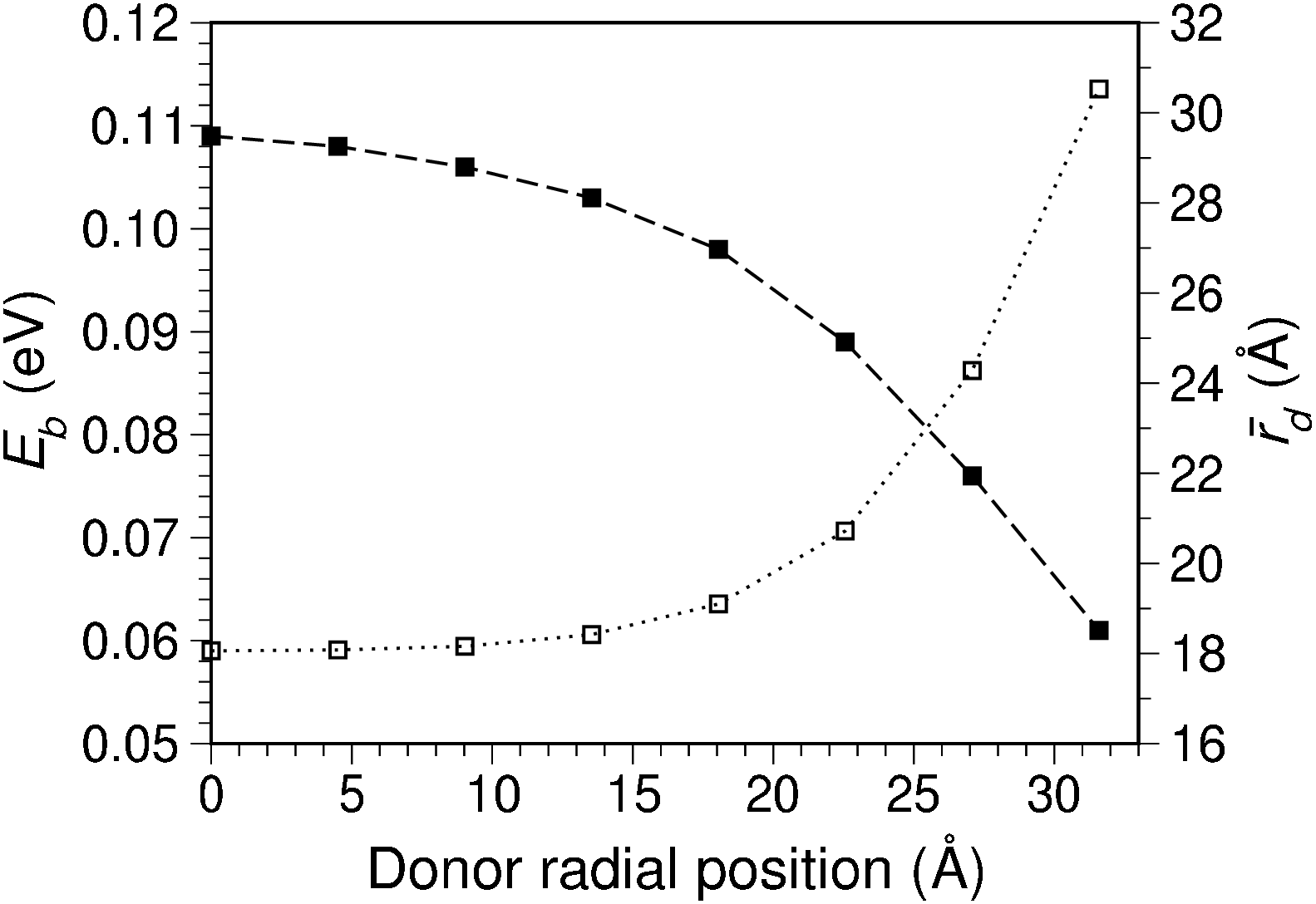}
\caption{\label{fig:fig9}Binding energy and average radius of the electron
ground state as a function of the donor position in a QD of radius 31.6~{\AA} (closed and open symbols, respectively).}
\end{figure}

On the other hand, in the case of donor impurity, the Bohr radius is comparable to the radius of the referred NC and the bound electron is spread over the whole NC. As a result, the position dependence of the binding energy and the average radius, as shown in Fig.~\ref{fig:fig9}, is quite smooth. The effects are important regardless of the O atom location. The closer the dopant gets to the crystal boundary, the smaller is the binding energy and the larger the average electron radius. This monotonously decreasing binding energy is comparable to that calculated in Ref.~\onlinecite{pssb}. From the experimental point of view, our results show that the acceptor-doped NC's are less sensitive to the impurity position than the donor-doped NC's.

%--------------------------------------------------------------------
%	Figura 8-Binding enery 
%--------------------------------------------------------------------

%--------------------------------------------------------------------
%	Figura 9-Binding energy
%--------------------------------------------------------------------

\newpage
%--------------------------------------------------------------------

%%%%%%%%%%%%%%%%%%%%%%%%%%%%%%%%%%%%%%%%%%%%%

\section{CONCLUDING REMARKS}

%%%%%%%%%%%%%%%%%%%%%%%%%%%%%%%%%%%%%%%%%%%%%
We have studied acceptor (Mg) and donor (O) states in GaN NC's doped with a single substitutional impurity within the $sp^3d^5s^{\ast}$ tight-binding model. The zinc-blende-structure crystallites are assumed spherical, ranging from 1 to 13.5 nm in diameter. The computed binding energy is highly enhanced with respect to the experimental bulk value when the dopant is placed at the center of the smallest QD's. For larger sizes it decreases following a scaling law that extrapolates to the bulk limit ($236$~meV for Mg and $33$~meV for O). The
degree of localization of the bound carriers is analyzed through their wave functions and average radii. The ground-state acceptor hole is mostly distributed over the nearest-neighbor anion shell. The donor electron, on the contrary, is much less localized. We also investigated off-center impurities in intermediate-size NC's ($R\sim32$~{\AA}). The acceptor binding energy is found to be barely dependent on the Mg impurity position unless it lies within a surface shell of $\sim4.5$~{\AA} thickness ($\sim$ the acceptor Bohr radius), where the ionization energy is drastically reduced. On the contrary, the donor binding energy gradually decreases as the O dopant approaches the surface (the donor Bohr radius $\sim28$~{\AA} is similar to the radius of the referred NC's). This difference arises from the larger spatial extension of the bound electron as compared to the bound hole.  Although we have focused on Mg and O impurities, our approach can be straightforwardly extended to other dopant species.
%%%%%%%%%%%%%%%%%%%%%%%%%%%%%%%%%%%%%%%%%%%%%

%\bibliography{bb}

\end{document}